\documentclass[prd,twocolumn,10pt,aps,longbibliography,nofootinbib]{revtex4-2}
\pdfoutput=1

\usepackage{subfig,tabularx,afterpage,xifthen}
\usepackage{amsmath,amsfonts,amssymb,mathtools}
\usepackage{slashed}
\usepackage{comment}
\usepackage{tikz}
\usepackage{graphicx}
\usepackage{lmodern}

\newcommand\Phat{P}
\newcommand\nbar{{\bar{n}}}
\newcommand\nb{{\bar{n}}}
\newcommand\n{n}

\newcommand{\overbar}[1]{\mkern 1.5mu\overline{\mkern-1.5mu#1\mkern-1.5mu}\mkern 1.5mu}
\newcommand\Deltabar{\overbar{\Delta}}

\newcommand\secref[1]{{Sec.\ \ref{#1}}}

\newcommand\qcd{\mathrm{QCD}}

\newcommand\psibar{\bar{\psi}}
\newcommand\chibar{\bar{\chi}}

\newcommand\ggamma{\gamma\kern-8.2pt \gamma}

\DeclareSymbolFont{matha}{OML}{txmi}{m}{it}
\DeclareMathSymbol{\varv}{\mathord}{matha}{118}

\def\OMIT#1{}

\def\eqn#1{Eq.\ (\ref{#1})}
\def\lqcd{\Lambda_{\rm QCD}}

\def\tmunu{g_\perp^{\mu\nu}}
\def\calj{{\cal J}}

\def\cj{C_J}

\begin{document}

\title{Factorization at subleading power in deep inelastic scattering in the $x\rightarrow 1$ limit}

\author{Michael Luke}
\email{luke@physics.utoronto.ca}
\author{Jyotirmoy Roy}
\email{jro1@physics.utoronto.ca}
\author{Aris Spourdalakis}
\email{aspourda@physics.utoronto.ca}

\affiliation{Department of Physics, University of Toronto, Toronto, Ontario M5S 1A7, Canada }
\begin{abstract}
We examine the endpoint region of inclusive deep inelastic scattering at next-to-leading power (NLP). Using a soft-collinear effective theory approach with no explicit soft or collinear modes, we discuss the factorization of the cross section at NLP and show that the overlap subtraction procedure introduced to eliminate double counting of degrees of freedom at leading power ensures that spurious endpoint divergences in the rate cancel at NLP at one loop. For this cancellation to occur at all renormalization scales a nontrivial relation between the anomalous dimensions of the leading and subleading operators is required, which is demonstrated to hold at one loop.
\end{abstract}

\maketitle

%
\section{Introduction}

Soft-collinear effective theory (SCET) \cite{Bauer:2000ew,Bauer:2000yr,Bauer:2001ct,Bauer:2001yt,Bauer:2002nz,Beneke:2002ph,Beneke:2002ni} is a well-established tool for studying hard scattering processes in QCD as an expansion in inverse powers of the hard scattering scale $Q$. While there has been much work on applications of SCET at leading power (LP), power corrections suppressed by inverse powers of $Q^2$ have proven more involved, in part due to the appearance of additional divergences which spoil naive factorization.  Recent work studying power corrections to various processes in SCET include beam thrust \cite{Moult:2018jjd}, Drell-Yan production near threshold \cite{Beneke:2018gvs,Beneke:2019oqx} and at small $q_T$ \cite{Inglis-Whalen:2021bea}, threshold Higgs production from gluon fusion \cite{Beneke:2019mua}, Higgs production and decay \cite{Bhattacharya:2018hmss}, the energy-energy correlator in $\mathcal{N}=4$ Supersymmetric Yang-Mills \cite{Moult:2019vou}, Higgs to diphoton decay \cite{Liu:2019oav,Liu:2020tzd,Liu:2020wbn} off-diagonal deep inelastic scattering \cite{Beneke:2020ibj}, gluon thrust \cite{Beneke:2022obx} and muon-electron backward scattering \cite{Bell:2022ott}. Power corrections have also been studied using non-EFT QCD techniques \cite{Kramer:1996iq,Penin:2015hel,Bonocore:2015esa,Bonocore:2016awd,Bahjat-Abbas:2019fqa, Cieri:2019hopc, vanbeekveld:2021hhv, Oleari:2020wvt, Boughezal:2020vjp,vanBeekveld:2021mxn}.

SCET has complications not found in more familiar effective field theories (EFT's) such as four-Fermi theory or heavy quark effective theory because it simultaneously describes particles with parametrically different momentum scaling.  In its most familiar formulations the relevant modes contributing to a given process (soft, collinear, ultrasoft, hard-collinear as well as others, depending on the scales of interest) are described by separate fields.  In a more recent formalism \cite{Goerke:2017ioi} it was argued that SCET is more simply written as a theory of separate sectors, defined such that the invariant mass of pairs of particles in different sectors is of order $Q^2$, but the invariant mass of pairs of particles in the same sector is parametrically smaller than $Q^2$. Particles in different sectors are described by different fields, but modes of a given particle in a single sector are described by the same field, as in QCD. In every formulation, however, spurious divergences arise in individual graphs in SCET because loop- and phase-space integrals integrate over all momenta, including momenta which violate the power counting of the corresponding field, giving unphysical contributions. As was demonstrated many years ago \cite{Manohar:2006nz}, matrix elements in SCET are only well-defined if an appropriate subtraction procedure has been implemented to remove this double counting between different modes or sectors. Since the EFT by construction must reproduce the physics of QCD these divergences must cancel in physical observables once the appropriate subtractions have been made; however, this is not always simple to demonstrate.

Endpoint divergences in particular are unphysical divergences arising from convolutions of Wilson coefficients and operators in SCET, and lead to an apparent violation of factorization. The appearance of these endpoint divergences is a common feature at NLP and has been recently studied in context of various processes \cite{Liu:2019oav,Liu:2020tzd,Liu:2020wbn,Beneke:2020ibj,Beneke:2022obx}. While individual terms in the factorization formula are divergent, it was demonstrated in these works that the factorized physical quantities remain finite since the divergences cancel between different terms in the endpoint region. This property was exploited to rearrange and rewrite the individual terms in a ``refactorized" form.
 
In this work, we examine next-to-leading power (NLP) corrections to deep inelastic scattering (DIS) \cite{Collins:1981uw,Ellis:1996mzs} in the endpoint limit using the formalism introduced in \cite{Goerke:2017ioi}. DIS in this limit was one of the first and simplest processes studied in SCET \cite{Manohar:2003vb,Becher:2006mr,Idilbi:2007ff}, and provides a simple example of a process with endpoint divergences at NLP.  The cross section is a function of the invariant mass $-q^2\equiv Q^2\gg\lqcd^2$ of the off shell photon and the dimensionless variable $x\equiv \frac{-q^2}{2P\cdot q}$, where $P^\mu$ is the four-momentum of the incoming proton and $q^\mu$ is the four-momentum of the photon. The cross section is well known to factorize into a hard scattering amplitude, depending on $Q^2$, and nonperturbative parton distribution functions (PDFs), which depend on the details of low-energy QCD. Large logarithms of $Q^2/\mu^2$ in the cross section, where $\mu\sim \lqcd$, may be resummed by evolving the PDFs using the DGLAP equations.
In the endpoint region where $x \rightarrow 1$, additional large logarithmic corrections appear in the perturbative expansion of the hard scattering amplitude which must be resummed to obtain a reliable calculation of the cross section.  These arise because in this limit the invariant mass of the final state, $p_F^2\sim Q^2(1-x)$, is parametrically smaller than the hard scattering scale $Q^2$.

The power corrections studied here correspond to terms suppressed by a single power of $1-x$ in the cross section relative to the leading terms. In \cite{Inglis-Whalen:2021bea}, it was shown that the overlap subtraction procedure introduced to correctly reproduce the infrared behavior of QCD in Drell-Yan (DY) scattering led automatically to the cancellation of rapidity divergences in the EFT, but that at NLP the cancellations typically involved linear combinations of multiple operators, including power suppressed subtractions of the leading-order operator. In the DY process the corresponding divergences were rapidity divergences, which require an additional regulator \cite{Becher:2010tm,Chiu:2011qc,Chiu:2012ir,Moult:2019vou,Ebert:2018gsn}. Here we show that the same overlap subtraction also ensures that the DIS rate obtained from SCET is free of endpoint divergences, but in this case the divergences are regulated in dimensional regularization. Again, the required cancellation occurs between separate operators, giving a nontrivial relation between the anomalous dimension of the leading and subleading operators which is demonstrated to hold at one loop. We defer an all-orders proof to a future work.

In \secref{Hard_Scale_Matching}, we review the scattering operators in SCET for DIS topology up to $O\left( 1/Q^{2}\right)$. We present the one-loop matrix elements of the product of operators required for inclusive DIS rate at NLP in \secref{NLP results} and show how the overlap procedure removes the endpoint singularities. In \secref{Resumm} we show that this cancellation of endpoint divergence holds at all scales and thus puts constraints on the anomalous dimension of subleading operators. Finally we present our conclusions in \secref{Conclusion}.

\section{Current to $O(1/Q^2)$}\label{Hard_Scale_Matching}

The incoming state in DIS consists of low invariant mass partons, $p_I^2\sim\lqcd^2$, while the outgoing state consists of partons with invariant mass $p_F^2\sim Q^2(1-x)$, with $p_I\cdot p_F\sim Q^2$. Thus, near $x=1$ they are described by different sectors in SCET which we denote by the lightlike vectors $n^\mu$ and $\nb^\mu$, where $n^2=\nb^2=0$ and $n\cdot\nb=2$. Partons in the incoming state are described by the $\nb$ sector, and particles in the outgoing state by the $n$ sector. Particles in the $n$ sector have momenta $p_n\cdot n\ll Q$, while $p_n\cdot\nb$ can be of order $Q$ (and vice versa for the $\nb$ sector). For simplicity, we work in a reference frame where $q^\mu$ has no components perpendicular to $n$ and $\nb$: $q^\mu=\nb \cdot q \frac{n^\mu}{2} +n \cdot q \frac{\nb^\mu}{2} $.

For simplicity we consider a single flavor of quark $\psi$. DIS is then mediated by the quark electromagnetic current
\begin{equation}
    J^\mu= \bar\psi\gamma^\mu \psi.
\end{equation}
At the hard scale $\mu \sim Q$, degrees of freedom with invariant mass of order $Q$ are integrated out of the theory, and QCD matrix elements are expanded in inverse powers of $Q$ and matched onto SCET. In the formalism used here \cite{Goerke:2017ioi}, each low-invariant mass sector of the theory is described by a different copy of QCD, with interactions between sectors occurring via Wilson lines in the external current.

Using the operator basis defined in \cite{Goerke:2017ioi, Goerke:2017lei,Inglis-Whalen:2021bea},
the current in the EFT is
\begin{equation} 
\begin{aligned} \label{eq:SSLO_expansion}
\calj^{\mu}(x) =   \sum_i \frac{1}{Q^{[i]}} C_2^{(i)}\left(\mu\right) O_2^{(i) \mu}(x,\mu)
\end{aligned} \end{equation}
where the $C_2^{(i)}$'s are Wilson coefficients and the $O_2^{(i)}$'s are operators in the EFT. Operators in SCET  are conveniently expressed in terms of the gauge invariant building blocks \cite{Kolodrubetz:2016slbb}
\begin{equation} 
\begin{aligned} \label{eq:building_blocks}
\chibar_n(x) &= \psibar_n(x) \overline{W}_n(x)\Phat_\nbar \\
\chi_\nbar(x) &= \Phat_\nbar W^\dagger_\nbar(x)\psi_\nbar(x) \\
\mathcal{B}_\n^{\mu_1 \cdots \mu_N}(x)&= \overline{W}_\n^\dagger(x) iD_\n^{\mu_1}(x) \cdots iD_\n^{\mu_N}(x) \overline{W}_\n(x) \\
\mathcal{B}_{\nb}^{\dagger\mu_1 \cdots \mu_N}(x) &= (-1)^N\, W_{\nb}^\dagger(x) i\overleftarrow{D}_{\nb}^{\mu_1}(x) \cdots i\overleftarrow{D}_{\nb}^{\mu_N}(x) W_{\nb}(x)
\end{aligned} 
\end{equation}
where $i D_{n}^{\mu}(x)=i\partial^\mu + gA_n^\mu(x)$ and $i\overleftarrow{D}_{\nb}^{\mu}(x) = i\overleftarrow{\partial}^\mu - gA_{\nb}^\mu(x)$ are the usual covariant derivatives in each sector, and $P_n=\slashed{n}\slashed{\nb}/4$, $P_\nb=\slashed{\nb}\slashed{n}/4$ are projection operators acting on the four-component spinors $\psi$. The outgoing Wilson lines in the $\nb$ sector and the incoming Wilson lines in the $n$ sector are defined respectively as

\begin{equation}
\begin{aligned} \label{eq:Wline_defns}
W_\nb^\dagger(x) &= \mathcal{P} \exp \left( ig_{s} \int_{0}^\infty \!\!\! ds\,\n\cdot A^a_{\nb}(x+\n s) T^a e^{-s 0^{+}} \right) \\
\overline{W}_\n(x) &= \mathcal{P} \exp \left( ig_{s} \int_{-\infty}^{0} \!\!\! ds\,\bar n\cdot A^a_{n}(x+\nb s) T^a e^{ s 0^{+}} \right)
\end{aligned} 
\end{equation}
Note that the subscript on a Wilson line corresponds to the sector with which it interacts rather than its direction.

The operators in the effective theory for DIS topology have already been derived in \cite{Goerke:2017ioi} up to $O(1/Q)$ using spinor-helicity techniques, and related operators relevant for Drell-Yan scattering and dijet production up to $O(1/Q^2)$ in \cite{Inglis-Whalen:2021bea,Goerke:2017lei}. Since we will not be using the helicity basis, it is instructive to write down the expanded amplitude and the corresponding SCET operators up to $O(1/Q^2)$ in terms of the usual Dirac matrices.
\begin{figure}[tbh!]
\begin{centering}
\includegraphics[width=0.35\textwidth]{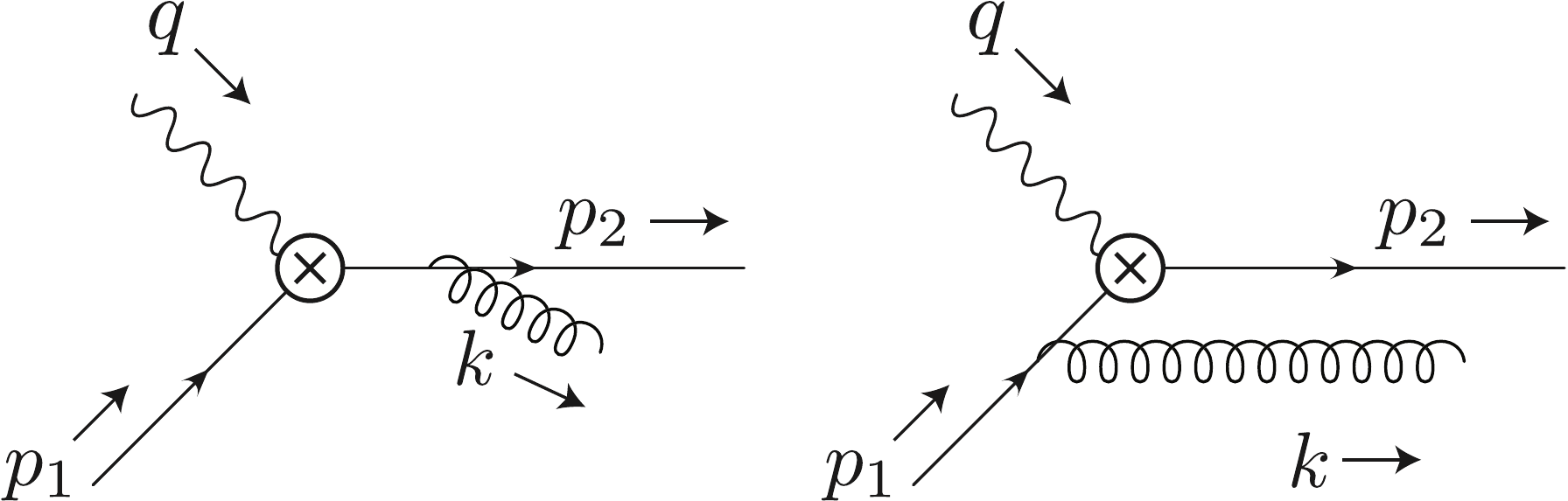}
\caption{One-gluon matrix element of $J^\mu$ in QCD.}
\label{fig:currentQCD}
\end{centering}
\end{figure}

The DIS amplitude to produce an additional gluon in the final state in Fig.~\ref{fig:currentQCD} is reproduced in SCET by different operators depending on whether the outgoing quark and gluon are in the $n$ or $\nb$ sector. For DIS near $x=1$ the appropriate assignment is the $n$ sector for both final-state particles. Expanding the QCD amplitude in powers of the small quantities $\frac{\nb \cdot p_1}{Q}$, $\frac{n\cdot p_2}{Q}$ and $\frac {n\cdot k}{Q}$ then gives
\begin{equation}
    \begin{aligned}\label{eq:qcdamplitude}
      i \mathcal{A}^{\mu} =& -g_{s} T^{a} \bar{u}(p_2) \left[ \frac{2p_{2}^{\alpha}+\gamma^{\alpha}\slashed{k}}{2p_{2} \cdot k} P_{\nbar}\gamma^{\mu}P_{\nbar} -P_{\nbar}\gamma^{\mu}P_{\nbar} \frac{\nbar^{\alpha}}{\nbar \cdot k} \right. \\
      &  +\frac{1}{Q} \left( \Deltabar^{\alpha \beta}(k) \gamma^{\perp}_{\beta} \frac{\slashed{\nbar}}{2}\gamma^{\mu}P_{\nbar}+P_{\nbar}\gamma^{\mu}\frac{\slashed{n}}{2}\gamma^{\perp}_{\beta} \Deltabar^{\alpha \beta}(k) \right)\\ 
      & +\frac{1}{Q^2} P_{\nbar}\gamma^{\mu}P_{\nbar} \gamma^{\perp}_{\beta} \gamma^{\perp}_{\gamma} k^{\beta} \Deltabar^{\alpha \gamma}(k) \left( 1+\frac{\nb \cdot p_2}{\nb \cdot k} \right)  \\
      & \left. - \frac{1}{Q^2}\gamma^{\perp}_{\beta} \frac{\slashed{\nbar}}{2}\gamma^{\mu} \frac{\slashed{n}}{2} \gamma^{\perp}_{\gamma} k^{\beta} \Deltabar^{\alpha \gamma}(k) \left(1+ \frac{\nb \cdot k}{\nb \cdot p_2} \right) \right] \\
      &\times u(p_1) \varepsilon^{*}_{\alpha}(k)+\dots \ 
    \end{aligned}
\end{equation}
where $\Deltabar^{\alpha \beta}(k)=g^{\alpha \beta}-\frac{\nb^{\alpha}k^{\beta}}{\nb \cdot k}$, and we have chosen to work in a reference frame in which the total perpendicular momentum in both the $n$ and $\nb$ sectors vanishes (this is in contrast with Drell-Yan production \cite{Inglis-Whalen:2021bea}, where such a choice is not generally possible). 

The LP terms in \eqn{eq:qcdamplitude} are reproduced by the SCET operator
\begin{equation}
    O_2^{(0)\mu}(x)=\chibar_n(x) \gamma^\mu \chi_\nbar(x)
\end{equation}
while the $1/Q$ term is given by the operator
\begin{equation}\label{eq:1A}
\begin{aligned}
  O_{2}^{(1A)\mu}(x,\hat{t})=&-\chibar_n(x) \mathcal{B}_n^{\alpha}(x+\nb t)\\&\times  \left( \gamma_{\alpha}^{\perp}\frac{\slashed{\nb}}{2}\gamma^{\mu}+\gamma^{\mu}\frac{\slashed{n}}{2}\gamma_{\alpha}^{\perp} \right) \chi_\nbar(x)
  \end{aligned}
\end{equation}
(note that the relative sign between the two terms is opposite to that in DY).
Here we have introduced the continuous shift parameter $\hat{t} \equiv (\nb \cdot q ) t$ for shift in the $n$ sector and $\hat{t} \equiv (n \cdot q) t$ for shift in the $\nb$ sector which parametrizes the separation of fields along the light cones. Finally the $1/Q^2$ terms are reproduced by the two operators
\begin{equation}\label{eq:2A}
    \begin{aligned}
      O_{2}^{(2A_1)\mu}(x,\hat{t})=& 2\pi i \theta(\hat{t})\\& \otimes \chibar_n(x) \mathcal{B}_n^{\alpha \beta}(x+\nb t)  \gamma^{\mu} \gamma_{\alpha}^{\perp} \gamma_{\beta}^{\perp}  \chi_\nbar(x) \\
      O_{2}^{(2A_2)\mu}(x,\hat{t})=& -2\pi i \theta(\hat{t}) \\&\otimes \chibar_n(x+\nb t) \mathcal{B}_n^{\alpha \beta}(x)  \gamma_{\alpha}^{\perp} \frac{\slashed{\nb}}{2} \gamma^{\mu} \frac{\slashed{n}}{2} \gamma_{\beta}^{\perp}  \chi_\nbar(x)
    \end{aligned}
\end{equation}
where we have defined the convolutions in $\hat t$ by
\begin{equation} \begin{aligned} \label{eq:convolution}
f(\hat{t}) \otimes g(\hat{t}) = \int\! \frac{dx\,dy}{2\pi}\, f(x) g(y) \, \delta(\hat{t}-x-y).
\end{aligned} 
\end{equation}
The QCD amplitude with the final-state quark in the $n$ sector and the gluon in the $\nb$ sector is reproduced by the operators 
\begin{equation}\label{eq:2B}
    \begin{aligned}
      O_{2}^{(1B)\mu}(x,\hat{t})=& -\chibar_n(x) \left( \gamma_{\alpha}^{\perp}\frac{\slashed{n}}{2} \gamma^\mu+ \gamma^{\mu}\frac{\slashed{\nb}}{2}\gamma_{\alpha}^{\perp} \right)\\
      &\times \mathcal{B}_{\nb}^{\dagger \alpha}(x-n t) \chi_\nbar(x) \\
      O_{2}^{(2B_1)\mu}(x,\hat{t})=& 2\pi i \theta(\hat{t}) \\ &\otimes \chibar_n(x)  \gamma_{\alpha}^{\perp} \gamma_{\beta}^{\perp} \gamma^{\mu}  \mathcal{B}_{\nb}^{\dagger \alpha \beta}(x-n t) \chi_\nbar(x) \\
      O_{2}^{(2B_2)\mu}(x,\hat{t})=& -2\pi i \theta(\hat{t}) \\ &\otimes \chibar_n(x)  \gamma_{\alpha}^{\perp} \frac{\slashed{n}}{2} \gamma^{\mu} \frac{\slashed{\nb}}{2}\gamma_{\beta}^{\perp}  \mathcal{B}_{\nb}^{\dagger \alpha \beta}(x) \chi_\nbar(x-nt)
\end{aligned}\end{equation}
while with the final-state quark in the $\nb$ sector and the gluon in the $n$ sector is reproduced by the operators
\begin{equation}\label{eq:2C}
    \begin{aligned}
      O_{2}^{(1C_1)\mu}(x,\hat{t})=& 2\pi i \theta(\hat{t})\\& \otimes \mathcal{B}_{n}^{\alpha c c'}(x) \chibar^{c}_{\nb}(x-n t)   \gamma_{\alpha}^{\perp}\frac{\slashed{n}}{2} \gamma^\mu \chi^{c'}_{\nb}(x) \\
      O_{2}^{(1C_2)\mu}(x,\hat{t})=& 2\pi i \theta(\hat{t})\\& \otimes \mathcal{B}_{n}^{\alpha c c'}(x) \chibar^{c}_{\nb}(x)   \gamma^\mu \frac{\slashed{n}}{2} \gamma_{\alpha}^{\perp}  \chi^{c'}_{\nb}(x-n t).
    \end{aligned}
\end{equation}
Near the endpoint of DIS the final state has a small invariant mass, so the final-state quark and gluon must be in the same sector, as particles are only described by separate sectors if their invariant mass is large. The operators (\ref{eq:2B}) and (\ref{eq:2C}) therefore do not contribute to our results, but we include them here for completeness. (More precisely, as will be discussed in the next section, if these operators are included their contributions will be completely canceled by the overlap subtraction procedure.)

With the above operator definitions, all matching coefficients are $C_2^{(i)}(\hat{t})=\delta(\hat{t})+O(g_{s})$. 
Following \cite{Hill:2004nlp}, we will be working with the Fourier-transformed operators
\begin{equation} \begin{aligned} \label{eq:fourier_conventions}
O_2^{(i)}(x,u)&= \int \! \frac{d\hat{t}}{2\pi} e^{-iu\hat{t}} 
O_2^{(i)}(x,\hat{t}\,) \\
C_2^{(i)}(x,u)&= \int  d\hat{t}\ e^{iu\hat{t}} \ C_2^{(i)}(x,\hat{t}\,).
\end{aligned} \end{equation}
The matching coefficients of the Fourier-transformed operators are then $C_2^{(i)}(u)=1+O(\alpha_s)$.

\section{Matching onto the PDF's}\label{NLP results}
The inclusive DIS rate is determined by matrix elements of the discontinuity of the \textit{T} product of electromagnetic currents in SCET,
\begin{equation}
    T^{\mu\nu}={\rm Disc}\,\frac1{2\pi} \int d^d x \ e^{-i q\cdot x} T\left[\calj^{\mu \dagger}(x) \calj^{\nu}(0) \right] .
\end{equation}
At a scale $\mu\sim Q\sqrt{1-x}$, the $n$-collinear sector is integrated out of the theory, and $T^{\mu\nu}$ is matched onto the bilocal quark-distribution operator $\phi(x)$
\begin{equation}\label{pdfmatching}
T^{\mu\nu} \to \int \frac{dw}{w}C^{\mu\nu}(w)\phi(-q^+/w)+\dots
\end{equation}
where
\begin{equation}\begin{aligned}\label{loPDF}
\phi(r^+) &= \frac{1}{4\pi}\int_{-\infty}^{\infty}dt \, e^{-i  r^+ t} \, \bar{\chi}_{\nb}(nt)\slashed{n}\chi_{\nb}(0)\\
&= \frac{1}{4\pi}\int_{-\infty}^{\infty}dt \, e^{-i  r^+ t} \, \bar{\psi}(nt)W(nt,0)\slashed{n}\psi(0),
\end{aligned}\end{equation}
and the ellipses denote higher-twist PDFs which we are not considering here. 
The quark PDF in a target $T$ with momentum $P$ is then given by the matrix element of $\phi(x)$ \cite{Collins:1981uw}
\begin{equation}
    f_{q/T}(x)=\langle T,P|\phi(x P^+)|T,P\rangle.
\end{equation}

The matching coefficient $C^{\mu\nu}$ is determined by taking matrix elements of \eqn{pdfmatching} between quark states at the matching scale $\mu\sim Q\sqrt{1-x}$. It is convenient to decompose the spin-averaged SCET matrix element into two Lorentz structures,
\begin{equation}
    \begin{aligned}
    \sum_{\rm spins}\langle p_\nb |  T^{\mu\nu}| p_\nb\rangle\equiv F_{T} \tmunu + F_{L} L^{\mu \nu}
    \end{aligned}
\end{equation}
where $|p_\nb\rangle$ denotes a quark state in the $\nb$ sector with momentum $p$, and
\begin{equation}
    \begin{aligned}
      g_\perp^{\mu\nu}\equiv &g^{\mu\nu}-\frac12(n^\mu \nb^\nu+\nb^\mu n^\nu),\\
      L^{\mu\nu}\equiv &\frac14( n^\mu+\nb^\mu)(n^\nu+\nb^\nu)
    \end{aligned}
\end{equation}
which satisfy $q_\mu g_\perp^{\mu\nu}=q_\nu g_\perp^{\mu\nu}=q_\mu L^{\mu\nu}=q_\nu L^{\mu\nu}=0$. The matrix elements are taken between states in the $\nb$ sector; since the $n$ sector is being integrated out of the theory, the corresponding matrix elements of $\phi(x)$ are the usual QCD PDFs.
We further define the spin-averaged matrix elements in $d=4-2\epsilon$ dimensions as
\begin{equation}
    \begin{aligned}
      F^{(i,j)}_{T}  \equiv &{\rm Disc} \ \frac{1}{2\pi (d-2) Q^{[i]+[j]}} \tmunu \\ &\times \sum_{\rm spins}\int d^d x \ e^{-i q\cdot x} \langle p_\nb | T \left[ O^{(i) \dagger}_{2 \mu}(x) O^{(j)}_{2 \nu}(0) \right] | p_\nb\rangle  \\
      F^{(i,j)}_{L}  \equiv &{\rm Disc} \ \frac{1}{2\pi Q^{[i]+[j]}} L^{\mu \nu}\\ &\times \sum_{\rm spins} \int d^d x \ e^{-i q\cdot x} \langle p_\nb | T \left[ O^{(i) \dagger}_{2 \mu}(x) O^{(j)}_{2 \nu}(0) \right] | p_\nb\rangle 
    \end{aligned}
\end{equation}
where the dimension of the operator $O_2^{(i)}$ is $[i]+3$.

As was demonstrated by Manohar and Stewart \cite{Manohar:2003vb}, matrix elements in SCET are only well-defined if an appropriate subtraction procedure has been implemented to remove double counting between different modes or sectors. In the formalism presented here, the unphysical $n$-collinear limit of the $\nb$ sector and the corresponding $\nb$ limit of the $n$ sector must be subtracted in loop diagrams and phase space integrals; the procedure was referred to as overlap subtraction. At LP, overlap subtraction is required to make SCET loop integrals well-defined and to correctly reproduce phase-space integrals \cite{Goerke:2017ioi}. At NLP the same subtraction procedure must be applied; however, it is more involved since at NLP one must include not only the leading terms from the $O(1/Q^2)$ results, but also the subleading overlap from the LP operators \cite{Inglis-Whalen:2021bea}.

First we review the LP calculation with the overlap subtraction presented in \cite{Goerke:2017ioi}.
At leading power, $F_L^{(0,0)}$ vanishes at $O(\alpha_s)$, while $F_T^{(0,0)}$ receives contributions at $O(\alpha_s)$ from both $n$ and $\nb$ sector gluons. Their separate contributions are
\begin{equation}\label{O2n}
    \begin{aligned}
      F_{T,n}^{(0,0)}= & \frac{\alpha_s C_F}{2 \pi} \left\lbrace -\frac{2}{\epsilon^2} \delta(1-y)\right.\\
      &\left.+\frac{1}{\epsilon} \left( \left( 2 \log \frac{Q^2}{\mu^2} -\frac{3}{2} \right) \delta (1-y) +\frac{2}{[1-y]_{+}} \right)   \right.\\
      &  - \left(  \log^{2} \frac{Q^2}{\mu^2}-\frac{3}{2} \log \frac{Q^2}{\mu^2} -\frac{\pi^2}{2} +\frac{7}{2} \right) \delta (1-y) \\
      & \left.- \left( 2\log \frac{Q^2}{\mu^2 y}  - \frac{3}{2} \right) \frac{1}{\left[  1-y \right]_{+}} - 2 \left[ \frac{\log (1-y)}{1-y} \right]_{+} \right\rbrace 
    \end{aligned}
\end{equation}
and
\begin{equation}\label{O2nb}
    \begin{aligned}
      F_{T,\nb}^{(0,0)} =& \frac{\alpha_s C_F}{2 \pi} \left\lbrace  -\frac{2}{\epsilon^2} \delta(1-y)\right.
      \\  &+\frac{1}{\epsilon} \left( 2 \log \frac{Q^2}{\mu^2} \, \delta (1-y) +\frac{1+y^2}{[1-y]_{+}} \right) \\
      &  -\left( \log^{2} \frac{Q^2}{\mu^2} -\frac{\pi^2}{2} \right) \delta (1-y) -\log \frac{Q^2}{\mu^2 y} \, \frac{1+y^2}{\left[  1-y \right]_{+}} \\
      &- \left.\left( 1+y^2 \right) \left[ \frac{\log (1-y)}{1-y} \right]_{+}  -\frac{1}{2} (3+y) \theta(1-y) \right\rbrace 
    \end{aligned}
\end{equation}
where $y \equiv -\frac{n \cdot q}{n \cdot p}$.

The overlap subtraction term is given by expanding the $\nb$ sector graph in the ``wrong" limit $k\cdot n\ll k\cdot\nb$ and taking the leading-order term, which subtracts the leading-order contribution from integrating over the unphysical region of phase space. This gives
\begin{equation}\label{O2ov}
    \begin{aligned}
     F_{T,\nb \rightarrow n}^{(0,0)} =& \text{Disc} \ \frac{g^2 C_F}{8 \pi (1-\epsilon) Q^2} \left(\frac{\mu^2 e^\gamma}{4\pi}\right)^\epsilon \tmunu \\
     &\times \int \frac{d^{4-2\epsilon}k}{(2\pi)^{4-2\epsilon}} \ \frac{2}{(-n \cdot k \ \nb \cdot k)} \\
     & \times \frac{\ \text{Tr} \left[ \slashed{p} P_{n} \gamma_{\mu} P_{n} \left( \slashed{p}+\slashed{q}-\slashed{k} \right) P_{\nbar} \gamma_{\nu} P_{\nbar} \right] }{\left[(p+q-k)^{2}+i0^{+} \right] \left(k^{2}+i0^{+} \right)} \\
     =& \frac{\alpha_s C_F}{2 \pi} \left\lbrace  -\frac{2}{\epsilon^2} \delta(1-y)\right.
     \\& +\frac{1}{\epsilon} \left( 2 \log \frac{Q^2}{\mu^2} \, \delta (1-y) +\frac{2}{[1-y]_{+}} \right)    \\
     & -\left( \log^{2} \frac{Q^2}{\mu^2} -\frac{\pi^2}{2} \right) \delta (1-y)  \\
     &\left.-2\log \frac{Q^2}{\mu^2 y} \frac{1}{\left[  1-y \right]_{+}} - 2 \left[ \frac{\log (1-y)}{1-y} \right]_{+} \right\rbrace 
    \end{aligned}
\end{equation}
and cancels the contribution of the $\nb$ sector graph in the $y\to 1$ limit.
Adding together the contributions from the two sectors and subtracting the overlap as well as the ultraviolet counterterm for $O_2^{(0)}$, gives the well-known result
\begin{equation}
\begin{aligned}
 F_{T}^{\rm{LP}}=& -\delta(1-y) + \frac{\alpha_s C_F}{2 \pi} \left\lbrace  \frac{1}{\epsilon} \left( \frac{3}{2}\delta (1-y) +\frac{1+y^2}{[1-y]_{+}} \right)\right.\\
 & -\left( \log^{2} \frac{Q^2}{\mu^2} -\frac{3}{2} \log \frac{Q^2}{\mu^2} -\frac{\pi^2}{2} + \frac{7}{2} \right) \delta (1-y)\\
 &-\left( \left( 1+y^2 \right) \log \frac{Q^2}{\mu^2 y}  -\frac{3}{2} \right) \frac{1}{\left[  1-y \right]_{+}} \\
 &\left.- \left( 1+y^2 \right) \left[ \frac{\log (1-y)}{1-y} \right]_{+}   -\frac{1}{2} (3+y) \theta(1-y) \right\rbrace \ .
\end{aligned}
\end{equation}
The remaining $1/\epsilon$ terms are infrared divergences which are reproduced by the perturbative matrix element of $\phi(x)$, so the final expression for the matching coefficient at LP takes the factorized form
\begin{equation}
    C^{\mu \nu}(w)=\left| C_{2}^{(0)}(\mu) \right|^{2}\cj^{(0,T)}(w) \tmunu + \dots
\end{equation}
where $\cj^{(0,T)}(w)$ is the leading power matching coefficient obtained from the finite parts of $F_T^{\rm{LP}}$,
\begin{equation}
\begin{aligned}
    \cj^{(0,T)}(w) = & -\delta(1-w) \\
    &- \frac{\alpha_s C_F}{2 \pi} \left\lbrace   \left( \log^{2}  \frac{Q^2}{\mu^2} -\frac{3}{2} \log \frac{Q^2}{\mu^2} \right.\right.
    \\&\left.\left.-\frac{\pi^2}{2} + \frac{7}{2} \right) \delta (1-w) \right. \\
 &  \left.  +\left( \left( 1+w^2 \right) \log \frac{Q^2}{\mu^2 w}  - \frac{3}{2} \right) \frac{1}{\left[  1-w \right]_{+}} \right.\\
 &\left.+ \left( 1+w^2 \right) \left[ \frac{\log (1-w)}{1-w} \right]_{+} \right. \\
 &\left.+\frac{1}{2} (3+w) \theta(1-w) \right\rbrace +O(\alpha_s^2)\ 
 \end{aligned}
\end{equation}
and we have pulled out the explicit factor of the hard matching coefficient $|C_2(\mu)|^2$ from the product of currents (\ref{pdfmatching}) to define $\cj^{0,T}(w)$.

At NLP, only the operators (\ref{eq:1A})--(\ref{eq:2A}) contribute at this order since those in (\ref{eq:2B}) and (\ref{eq:2C}) have the wrong power counting. (Equivalently, they may be included, but are immediately subtracted away via overlap subtraction, as will be shown explicitly at the end of this section.) Since each of the relevant NLP operators has vanishing 0-gluon matrix element, at one loop only $n$ sector gluons contribute to the matrix elements of these operators.  The only nonvanishing contribution to the transverse structure function comes from the matrix elements $F^{(0,2A_1)}_{T,n}$ and its Hermitian conjugate $F^{(2A_1,0)}_{T,n}$,  
\begin{equation}
    \begin{aligned}
     &F^{(0,2A_1)}_{T,n}= \text{Disc} \ \frac{g^2 C_F}{4 \pi (2-2\epsilon) Q^2} \left(\frac{\mu^2 e^\gamma}{4\pi}\right)^\epsilon\\& \times\tmunu \int \frac{d^{4-2\epsilon}k}{(2\pi)^{4-2\epsilon}} \ \delta \left( u- \frac{\nb \cdot k}{\nb \cdot q} \right)  \frac{1}{u} k^{\beta} \Deltabar^{\alpha \gamma}(k)\\& \times\frac{\text{Tr} \left[ \slashed{p} P_{n} \gamma_{\mu} P_{n} \left( \slashed{p}+\slashed{q} \right) \gamma_{\alpha} \left( \slashed{p}+\slashed{q}-\slashed{k} \right) P_{\nbar} \gamma_{\nu} P_{\nbar} \gamma_{\beta}^{\perp} \gamma_{\gamma}^{\perp} \right]}{ \left[(p+q)^{2}+i0^{+} \right] \left[(p+q-k)^{2}+i0^{+} \right] \left(k^{2}+i0^{+} \right)}\\
     &= \frac{\alpha_{s} C_F}{2 \pi} \left(\frac{\mu^2 e^\gamma}{Q^2}\right)^{\epsilon} \frac{1}{\Gamma(1-\epsilon)} \frac{\theta(1-y)}{(1-y)^{\epsilon}y^{1-\epsilon}}\\
     &\times\frac{(1-u)^{2-\epsilon} \theta(u) \theta(1-u)}{u^{1+\epsilon}}\ .
    \end{aligned}
\end{equation}
Note that this contribution has an endpoint divergence as $u\to 0$ which is regulated in dimensional regularization. Explicitly, using the tree-level matching condition $C_2^{(2A1)}(u)=1$, the tree-level rate is determined by the integral
\begin{equation}\label{eq:endpoint}
    \begin{aligned}
      \int du \, F_{T,n}^{(0,2A_1)}=&\frac{\alpha_s C_F}{2\pi}\frac{\theta(1-y)}y\\ &\times\left(-\frac1\epsilon+\log \frac{Q^2(1-y)}{\mu^2 y}-\frac{3}{2}\right)\ .
    \end{aligned}
\end{equation}
The $1/\epsilon$ divergence is spurious; it is not absorbed by an ultraviolet counterterm in the EFT, nor does it correspond to an infrared divergence in the Altarelli-Parisi splitting functions. Thus, it indicates that without the appropriate overlap subtraction, the EFT is not correctly reproducing the infrared divergences of QCD. 

Since there is no $\nb$ gluon contribution to $F_{T,n}^{(0,2A_1)}$, there is no overlap graph associated with the operator to cancel this divergence. However, as discussed in \cite{Inglis-Whalen:2021bea}, consistently canceling unphysical contributions from the wrong sector must be done order by order in $1/Q$. The overlap subtraction procedure for DIS involves expanding the graphs with $\nb$ sector gluons in the ``wrong" limit $k\cdot n\ll k\cdot\nb$. The leading-order term in this expansion gives the leading overlap (\ref{O2ov}). Expanding to NLP and then performing the loop integrals gives the subleading overlap 
\begin{equation}
    \begin{aligned}
      F^{(0,0),\text{NLP}}_{T,\nbar \rightarrow n} =& 2\,\text{Disc} \ \frac{g^2 C_F}{8 \pi (1-\epsilon) Q^2} \left(\frac{\mu^2 e^\gamma}{4\pi}\right)^\epsilon \tmunu\\
      &\times \int \frac{d^{4-2\epsilon}k}{(2\pi)^{4-2\epsilon}} \ \frac{2 (n \cdot k \ \nb \cdot q)}{(-n \cdot k \ \nb \cdot k)}  \\
      & \times \frac{\text{Tr} \left[ \slashed{p} P_{n} \gamma_{\mu} P_{n} \left( \slashed{p}+\slashed{q}-\slashed{k} \right) P_{\nbar} \gamma_{\nu} P_{\nbar} \right] }{\left[(p+q-k)^{2}+i0^{+} \right] \left(k^{2}+i0^{+} \right)} \\
      =& \frac{\alpha_s C_F}{\pi} \left(\frac{\mu^2 e^\gamma}{Q^2}\right)^\epsilon \\&\times \frac{2^{2\epsilon-1} \sqrt{\pi}\Gamma (2-\epsilon)}{(-\epsilon) \Gamma(1-\epsilon) \Gamma (\frac{3}{2}-\epsilon)} \frac{\theta(1-y)}{(1-y)^{\epsilon}y^{1-\epsilon}}\\
      =&\frac{\alpha_s C_F}{\pi}\frac{\theta(1-y)}y\left(-\frac1\epsilon+\log \frac{Q^2(1-y)}{\mu^2 y}-1\right)\\&+O(\epsilon)
    \end{aligned}
\end{equation}
which must be subtracted from the rate at NLP. This term cancels the spurious divergence in \eqn{eq:endpoint} (plus its Hermitian conjugate) and gives a finite NLP result for $F_T$
\begin{equation}\label{eq:ftscet}
    F_{T}^{\rm{NLP}}= -\frac{\alpha_s C_F}{2 \pi} \frac{\theta(1-y)}{y} \ .
\end{equation}
The longitudinal structure function gets a contribution only from the matrix element $F^{(1A,1A)}_{L,n}$,
\begin{equation}
    \begin{aligned}
      &F^{(1A,1A)}_{L,n} = \text{Disc} \ \frac{g^2 C_F}{4 \pi Q^2} \left(\frac{\mu^2 e^\gamma}{4\pi}\right)^\epsilon L^{\mu \nu}\\& \times\int \frac{d^{4-2\epsilon}k}{(2\pi)^{4-2\epsilon}} \ \delta \left( u- \frac{\nb \cdot k}{\nb \cdot q} \right) \delta \left( v- \frac{\nb \cdot k}{\nb \cdot q} \right) \Deltabar_{\gamma}^{\alpha}(k) \\
      & \times \Deltabar^{\gamma \beta}(k) \frac{\text{Tr} \left[ \slashed{p} \left( P_{n} \gamma_{\mu} \frac{\slashed{\nbar}}{2} \gamma_{\alpha}^{\perp} + \gamma_{\alpha}^{\perp} \frac{\slashed{n}}{2} \gamma_{\mu} P_{n} \right) \left( \slashed{p}+\slashed{q}-\slashed{k} \right)\right.} {\left[(p+q-k)^{2}+i0^{+} \right] \left( k^{2} +i 0^{+} \right)} \\
      &\times \left.\left( \gamma_{\beta}^{\perp} \frac{\slashed{\nbar}}{2} \gamma_{\nu} P_{\nbar} + P_{\nbar} \gamma_{\nu} \frac{\slashed{n}}{2} \gamma_{\beta}^{\perp} \right) \right]\\
      &= \frac{2 \alpha_s C_F} {\pi} \left(\frac{\mu^2 e^\gamma}{Q^2}\right)^{\epsilon} \frac{(1-\epsilon)}{\Gamma (1-\epsilon) }\frac{\theta(1-y)}{(1-y)^{\epsilon}y^{1-\epsilon}}\\ & \times(1-u) \theta(u)\theta (1-u) \delta (u-v)\ .
 \end{aligned}
\end{equation}
\OMIT{Similarly for the $\nbar$-sector, the matrix elements which contributes to the transverse functions are $F^{(0,2B_1)}_T$ and $F^{(2B_1,0)}_T$
\begin{equation}
    F^{(0,2B_1)}_T=\frac{\alpha_s C_F} {2 \pi} \left(\frac{\mu^2 e^\gamma}{Q^2}\right)^{\epsilon} \frac{(1-\epsilon+\epsilon^{2} u y)}{\Gamma (2-\epsilon)}\frac{y \theta(1-y)}{(1-y)^{2-\epsilon}} \frac{ \theta(u) \theta \left( \frac{1-y}{y}-u \right)}{u^{\epsilon}(1-y-u y)^{\epsilon}} 
\end{equation}}
There is no spurious $\nb$ sector contribution, so we obtain
\begin{equation}
    F_{L}^{\rm{NLP}}= \frac{\alpha_s C_F}{\pi} \frac{\theta (1-y)}{y} \ .
\end{equation}

We can check these results by comparing with the fixed-order results from QCD. The structure functions can be extracted from the QCD hadronic tensor
\begin{equation}
   \begin{aligned}
     W^{\mu \nu}_{\qcd}=& {\rm Disc} \ \frac{1}{2\pi} \sum_{\rm spins} \int d^d x \ e^{-i q\cdot x} \langle p |  T\left[J^{\mu}(x) J^{\nu}(0) \right] | p \rangle \\
     =& F_{T}^{\qcd} \tmunu + F_{L}^{\qcd} L^{\mu \nu}
   \end{aligned}
\end{equation}
Adding all the graphs we get for the transverse structure function,
\begin{equation}
    \begin{aligned}
      F_{T}^{\qcd} =& -\delta(1-y) + \frac{\alpha_s C_F}{2 \pi} \left\lbrace  \frac{1}{\epsilon} \left( \frac{3}{2}\delta (1-y) +\frac{1+y^2}{[1-y]_{+}} \right)\right.\\
      & - \left( \frac{3}{2} \log \frac{Q^2}{\mu^2} -\frac{\pi^2}{3} -\frac{9}{2} \right)\delta (1-y)  \\
      &  -\left( \left(1+y^2 \right) \log \frac{Q^2}{\mu^2 y} -\frac{3}{2} \right) \frac{1}{\left[  1-y \right]_{+}} \\
      &\left .- \left( 1+y^2 \right) \left[ \frac{\log (1-y)}{1-y} \right]_{+} -3 \theta(1-y) \right\rbrace \ .
    \end{aligned}
\end{equation}   
We also have the longitudinal structure function in QCD
\begin{equation}
    \begin{aligned}
     F_{L}^{\qcd} = \frac{\alpha_s C_F}{\pi} y\theta (1-y)
 \end{aligned}
\end{equation} 
The difference between QCD and SCET transverse structure function is given by
\begin{equation}
    F^\text{QCD}_{T}-F^\text{SCET}_{T}= -\frac{\alpha_{s}C_F}{4\pi} (1-y) \theta (1-y)+O\left( (1-y)^2 \right)
\end{equation}
while for the longitudinal structure function
\begin{equation}
    F^\text{QCD}_{L}-F^\text{SCET}_{L}=O\left( (1-y)^2 \right)\ .
\end{equation}
Since  terms of order $(1-y)$ are further subleading in the power counting, the effective theory properly reproduces QCD up to NLP. 


\OMIT{The final state in DIS near the endpoint is described in SCET by the $n$-sector; overlap subtraction is required because $\nb$-sector gluons are necessarily also emitted into the final state.} 
At NLP, SCET also contains the subleading operators in (\ref{eq:2B}) and (\ref{eq:2C}) which we have ignored since the power counting is incorrect in this region. However, it is straightforward to show that including such operators doesn't change the result, since their spurious contributions are automatically eliminated by overlap subtraction.

Consider the contribution of $O_2^{(2B_1)}$, which emits an $\nb$ gluon into the final state. Naively, this operator contributes to $F_T$ through $F^{(0,2B_1)}_{T,\nb}$ and its Hermitian conjugate $F^{(2B_1,0)}_{T,\nb}$,
\begin{equation}\begin{aligned}
    F^{(0,2B_1)}_{T,\nb}=&\frac{\alpha_s C_F} {2 \pi} \left(\frac{\mu^2 e^\gamma}{Q^2}\right)^{\epsilon}\frac{(1-\epsilon+\epsilon^{2} u y)}{\Gamma (2-\epsilon)}\\ 
    &\times \frac{y \theta(1-y)}{(1-y)^{2-\epsilon}} \frac{ \theta(u) \theta \left( \frac{1-y}{y}-u \right)}{u^{\epsilon}(1-y-u y)^{\epsilon}}\ . 
\end{aligned}\end{equation}
and integrating over $u$ gives the contribution
\begin{equation}
    \int du \, F_{T,\nb}^{(0,2B_1)}=\frac{\alpha_s C_F} {2 \pi} \frac1{1-y}+O(\epsilon)
\end{equation}
which is leading order in $(1-y)$ and so violates power counting.
However, we must subtract the $\nb\to n$ limit of this graph,
\begin{equation}\begin{aligned}
      F^{(0,2B_1)}_{T,\nbar \rightarrow n}=&\frac{\alpha_s C_F} {2 \pi} \left(\frac{\mu^2 e^\gamma}{Q^2}\right)^{\epsilon} \frac{1}{\Gamma (1-\epsilon)}\\ &\times\frac{y \theta(1-y)}{(1-y)^{2-\epsilon}} \frac{ \theta(u) \theta \left( \frac{1-y}{y}-u \right)}{u^{\epsilon}(1-y-u y)^{\epsilon}} 
\end{aligned}\end{equation}
which, integrating this over $u$, gives the same contribution
\begin{equation}
\int du \, F_{T,\bar n\to n}^{(0,2B_1)}=\frac{\alpha_s C_F} {2 \pi} \frac1{1-y}+O(\epsilon)
\end{equation}
so $O(1/Q^2)$ contribution from the $\bar n$ sector operators is entirely removed by the overlap subtraction. Thus, the spurious contribution from this operator vanishes.

Similarly, the $\nb$ sector contribution to the longitudinal function is given by 
\begin{equation}
\begin{aligned}
    F^{(1B,1B)}_{L,\nb}=&\frac{2 \alpha_s C_F} {\pi} \left(\frac{\mu^2 e^\gamma}{Q^2}\right)^{\epsilon} \frac{(1-\epsilon)}{\Gamma (1-\epsilon)}\frac{y \theta(1-y)}{(1-y)^{2-\epsilon}} \\
    &\times\frac{u^{1-\epsilon} \theta(u) \theta \left( \frac{1-y}{y}-u \right)}{(1-y-u y)^{\epsilon}} \delta (u-v)
    \end{aligned}
\end{equation}
and is entirely canceled by its overlap,
\begin{equation}
\begin{aligned}
   F^{(1B,1B)}_{L,\nbar \rightarrow n}=& \frac{2 \alpha_s C_F} {\pi} \left(\frac{\mu^2 e^\gamma}{Q^2}\right)^{\epsilon} \frac{(1-\epsilon)}{\Gamma (1-\epsilon)}\frac{y \theta(1-y)}{(1-y)^{2-\epsilon}}\\
   &\times\frac{u^{1-\epsilon} \theta(u) \theta \left( \frac{1-y}{y}-u \right)}{(1-y-u y)^{\epsilon}} \delta (u-v)\ .
   \end{aligned}
   \end{equation}
It is straightforward to show that a similar cancellation occurs for the operators in (\ref{eq:2C}) where the final-state quark is in the $\nb$ sector and the gluon is in the $n$ sector.

\section{Renormalization Group Improvement} \label{Resumm}
 
To sum logarithms of $1-x$, SCET must be evolved from the hard scale $\mu=Q$ to the scale $\mu\sim Q\sqrt{1-x}$ where it is matched onto the PDF's. Since the cancellation of endpoint divergences depends on the interplay between the subleading overlap of $O_2^{(0)}$ and the matrix element of $O_2^{(2A_1)}$, this cancellation must occur independent of scale. This is straightforward to check at one loop.
   
The one-loop anomalous dimension of $O_2^{(2A_1)}$ was calculated in \cite{Goerke:2017lei}\footnote{The definition of $O_2^{(2A_1)}$ used here differs from that in \cite{Goerke:2017lei} by a factor of $u/v$, in addition to being defined for $q^2<0$ relevant for DIS, rather than $q^2>0$ for DY.},
\begin{equation}\label{RGE}
    \mu{d\over d\mu} O_2^{(2A_1)}(u,\mu)=-\int du\ \gamma_2^{(2A_1)}(u,v) O_2^{(2A_1)}(v,\mu)
\end{equation}
 where 
 \begin{equation}
    \begin{aligned}\label{eq:anomdim}
      \gamma^{(2A_1)}_{2}&(u,v) = \frac{\alpha_s}{\pi}\left( \delta (u-v) \left \lbrace C_F \left( \log \frac{Q^2}{\mu^2}-\frac{3}{2}+ \log \bar{v} \right)\right.\right.\\
      &\left.+\frac{C_A}{2} \left( \frac52+ \log \frac{v}{\bar{v}} \right) \right \rbrace \\
      & +\left( C_F-\frac{C_A}{2} \right) \left \lbrace \frac{\bar{u}^2}{u}\theta (u+v-1)\right. \\
      &\left.+\frac{v}{\bar{v}^2}(\bar{u}\bar{v}+\bar{u}+\bar{v}-1)\theta(1-u-v) \right\rbrace \\
      & -\left.\frac{C_A}{2u\bar{v}^2} \left \lbrace\vphantom{\frac{\theta(a)}{b}_c} v \bar{u}^2(1+\bar{v})\theta(u-v)+u \bar{v}^{2}(1+\bar{u}) \theta(v-u)\right.\right. \\
      &+\left.\left.\left[ v\bar{u}^{2} \frac{\theta(u-v)}{u-v} +u\bar{v}^{2} \frac{\theta(v-u)}{v-u}  \right]_{+} \right\rbrace\right),
    \end{aligned}
\end{equation}
and $\bar u=1-u$, $\bar v=1-v$. $\gamma^{(2A_1)}_{2}(u,v)$ is a nontrivial function of $u$ and $v$, so the solution to the renormalization group equation (\ref{RGE}) mixes operators with different values of $u$. However, at $O(\alpha_s)$ the current matches onto the linear combination 
\begin{equation}
     \begin{aligned}\label{eq:intoperators}
       \overline{O}_{2}^{(2A_1)}(\mu) \equiv \int_0^1 du \ O_{2}^{(2A_1)}(u,\mu)\ ,
     \end{aligned}
 \end{equation}
and this linear combination is multiplicatively renormalized. Explicitly, 
 \begin{equation}
    \begin{aligned}\label{eq:int_anomalousdimension}
      \mu \frac{d}{d\mu} \overline{O}_{2}^{(2A_1)}&(\mu) = \int_0^1 du \ \mu \frac{d}{d\mu} O_{2}^{(2A_1)}(u,\mu) \\
      =& - \int_0^1 du \ \int_0^1 dv \ \gamma^{(2A_1)}_{2}(u,v)  O_{2}^{(2A_1)}(v,\mu) \\
      =& -\int_0^1 dv \ O_{2}^{(2A_1)}(v,\mu) \int_0^1 du \ \gamma^{(2A_1)}_{2}(u,v).
      \end{aligned}
      \end{equation}
Integrating the expression in \eqn{eq:anomdim} gives, as required,
\begin{equation}
     \int_0^1 du \ \gamma^{(2A_1)}_{2}(u,v)=\gamma^{(0)}_2
\end{equation}
where 
\begin{equation}
    \gamma^{(0)}_{2} = \frac{\alpha_s C_F}{\pi} \left( \log \frac{Q^2}{\mu^2}-\frac{3}{2} \right)
\end{equation}
is the anomalous dimension of $O_2^{(0)}$. Thus, we have
\begin{equation}\label{eq:LOanom}
  \mu \frac{d}{d\mu} \overline{O}_{2}^{(2A_1)}(\mu) =-\gamma^{(0)}_{2} \overline{O}_{2}^{(2A_1)}(\mu)
\end{equation}
and so the terms have the same renormalization group evolution, at least at one loop. The cancellation of endpoint divergences therefore does not depend on the scale $\mu$, as is necessary for SCET to correctly reproduce the infrared physics of QCD. Consistency of the theory requires that this remains true at all orders, but a proof of this is beyond the scope of this paper.

This enables the calculation of \eqn{eq:ftscet} at any scale $\mu$ which is free of spurious divergences:
\begin{equation}
    F_{T}^{\rm{NLP}}(\mu) = -\frac{\alpha_s C_F}{2 \pi} \left| C_{2}^{(0)}(\mu)\right|^{2} \frac{\theta(1-y)}{y}\ .
\end{equation}
where $C_2^{(0)}(\mu)$ is evaluated with the leading-order anomalous dimension (\ref{eq:LOanom}).
In contrast, the operators contributing to the longitudinal structure do not have a simple renormalization group running and thus requires the explicit solution of the RGE in terms of the label $u$ to write the resummed form for $F_L$.

The matching coefficient onto the PDF therefore can be written in a factorized form up to NLP
\begin{equation}
    \begin{aligned}
      C^{\mu \nu}(w) =& \left|C_2^{(0)}(\mu)\right|^2\left[ \cj^{(0,T)}(w) +\cj^{(2,T)}(w) \right] \tmunu \\
      & + \int du \, dv  \, C_{2}^{(1A) \dagger}(u,\mu) C_{2}^{(1A)}(v,\mu) \\ &\times \cj^{(2,L)}(u,v,w) L^{\mu \nu}+\dots \ .
    \end{aligned}
\end{equation}
where $\cj^{(2,T)}(w)$ and $\cj^{(2,L)}(w)$ are the subleading matching coefficients for the transverse and longitudinal functions respectively,
\begin{equation}
    \begin{aligned}
      \cj^{(2,T)}(w) =& -\frac{\alpha_s C_F}{2 \pi} \frac{\theta(1-w)}{w}+O(\alpha_s^2)\\
      \cj^{(2,L)}(u,v,w) =& \frac{2 \alpha_s C_F} {\pi}\frac{\theta(1-w)}{w}(1-u)\\&\times   \theta(u)\theta (1-u) \delta (u-v)+O(\alpha_s^2)\ .
    \end{aligned}
\end{equation}

\section{Conclusions}\label{Conclusion}
We have calculated the leading-order power corrections to DIS near the endpoint in SCET. This is a particularly simple SCET calculation, but illustrates the point that spurious divergences which appear from the contributions of individual operators to the rate at NLP automatically cancel when the double-counting between sectors is correctly subtracted. At NLP this cancellation is more subtle than at LP because the cancellation in general only occurs after all operators contributing to a process and the corresponding double-counting have been included, and so requires nontrivial relations between the anomalous dimensions of different operators.  We note that a similar set of relations between anomalous dimensions of operators in the more standard (mode-based) treatment of SCET will also have to hold, since at subleading order the endpoint divergences cancel between different operators describing different modes.

While our calculations are only performed at the one-loop level, we expect the structure of cancellations appearing in this work to valid to all orders in $\alpha_s$, since this is required by consistency of the EFT and the requirement that it reproduce the infrared physics of QCD to all orders. In particular, the relation between the anomalous dimensions of $O_2^{(0)}(\mu)$ and $O_2^{(2A_1)}(\mu)$ should hold to all orders in $\alpha_s$. Work on this is in progress.

\bigskip 
\section*{Acknowledgments}

We thank Matt Inglis-Whalen for discussions about the running of the NLP operators, and Jaipratap Grewal and Aneesh Manohar for useful comments on the manuscript. This work was supported in part by the Natural Science and Engineering Research Council of Canada. 


\bibliographystyle{apsrev4-1}
\bibliography{nlp_DIS}
\end{document}